\documentclass[pra,aps,twocolumn]{revtex4-2}
\usepackage{graphicx,amsmath,color,soul}
\newcommand{\PT}{\mathcal{PT}}

\begin{document}

\title{Variational approach to study solitary waves in $\mathcal{PT}$-symmetric nonlinear couplers}

\author{Ambaresh Sahoo}
\email{ambareshs@iitg.ac.in}
\affiliation{Department of Physics, Indian Institute of Technology Guwahati, Assam 781039, India}

\author{Amarendra K. Sarma}
\email{aksarma@iitg.ac.in}
\affiliation{Department of Physics, Indian Institute of Technology Guwahati, Assam 781039, India}

\begin{abstract}
We theoretically investigate the solitary waves and their switching dynamics in a $\mathcal{PT}$-symmetric directional fiber coupler, exhibiting Kerr nonlinearity, by developing a variational analysis. We analyze the fundamental switching characteristics of the $\mathcal{PT}$-symmetric Kerr coupler in the picosecond timescale by considering the coupled-mode equation for the unperturbed nonlinear Schr\"{o}dinger equation, which we compare to its conventional counterpart. The impacts of higher-order perturbations (intrapulse Raman scattering, self-steepening, and third-order dispersion) are investigated in detail in the femtosecond timescale. In all cases, the variational method successfully predicts each of the numerically observed switching characteristics. Our semianalytical treatment has the potential to provide physical insights into complex switching dynamics in various nonlinear coupler configurations from different areas of physics.
\end{abstract}

\maketitle

\section{Introduction}
Non-Hermitian physics centered around $\PT$ symmetry has become a topic of intense research owing to numerous possible applications and for probing fundamental physics \cite{El-Ganainy18,Christodoulides}. It is amazing to note that while the idea of $\PT$ symmetry was first put forward in the context of Quantum mechanics \cite{Bender98, Bender99, Bender02}, it got validated in optical coupler settings \cite{Ruter10}. This is done by making an analogy between the Schr\"{o}dinger equation, with a complex potential, and the so-called Helmholz equation in optics, with a complex refractive index profile \cite{El-Ganainy07}. The idea of $\PT$ symmetry received experimental demonstrations in other optical settings as well \cite{Makris08,Guo09,Regensburger12}. Since the landmark experiment \cite{Ruter10} on the $\PT$-symmetric coupler was carried out, the area virtually got exploded with research from various perspectives. In particular, nonlinear $\PT$-symmetric coupler and its variants have been studied quite extensively \cite{Ramezani10, Sukhorukov10, He15, Yang18}. In this context, various solitary wave solutions to the $\PT$-symmetric coupled nonlinear Schr\"{o}dinger equations (NLSEs), that describe pulse propagation in a nonlinear $\PT$-symmetric coupler has been investigated by many research groups. The existence of stable vector solitons in the $\PT$-symmetric nonlinear coupler or similar systems with equal gain and loss or dissipation in the two cores or channels has been predicted numerically in several works \cite{Alexeeva12, Burlak13, Bludov13}. Again, stable small amplitude breathers have been theoretically demonstrated in such systems \cite{Barashenkov12}. 

Recently, soliton steering in nonlinear $\mathcal{PT}$-coupler has been thoroughly investigated, and significant improvement in critical power of switching over that of the conventional coupler is observed \cite{Govindarajan19, Sahoo21}. These studies may invoke renewed interests in nonlinear couplers as soliton switching devices. It should be noted that the aforementioned investigations are carried out only numerically. Admittedly, it is not possible to have exact analytical solutions to the $\PT$-symmetric coupled NLSEs. However, it is observed that even the semi-analytical methods like variational approach is rarely employed to study these systems. In the past, the variational approach has been successfully applied to study  NLSE to obtain useful physical insights \cite{Bondeson79, Anderson83}. The variational method has also proven to be effective in studying dissipative soliton dynamics in a variety of nonconservative systems \cite{Kaup95, Cerda98, Anderson99, Sahoo17, Sahoo19}. It is worthwhile to mention that the variational method is used by some authors to study solitary waves in the NLSE with complex potentials \cite{Mertens16}. In a recent study, we have carried out a variational study of the $\PT$-symmetric coupled NLSEs in the context of saturable nonlinearity \cite{Sahoo22}. In this work, we report a detailed variational study of coupled NLSEs in the context of soliton switching in a $\PT$-symmetric Kerr nonlinear coupler in the presence of higher-order perturbations. The results obtained are compared with numerical simulations, and found to agree considerably. 
%
\begin{figure}[t]
\centering
\begin{center}
\includegraphics[width=0.44\textwidth]{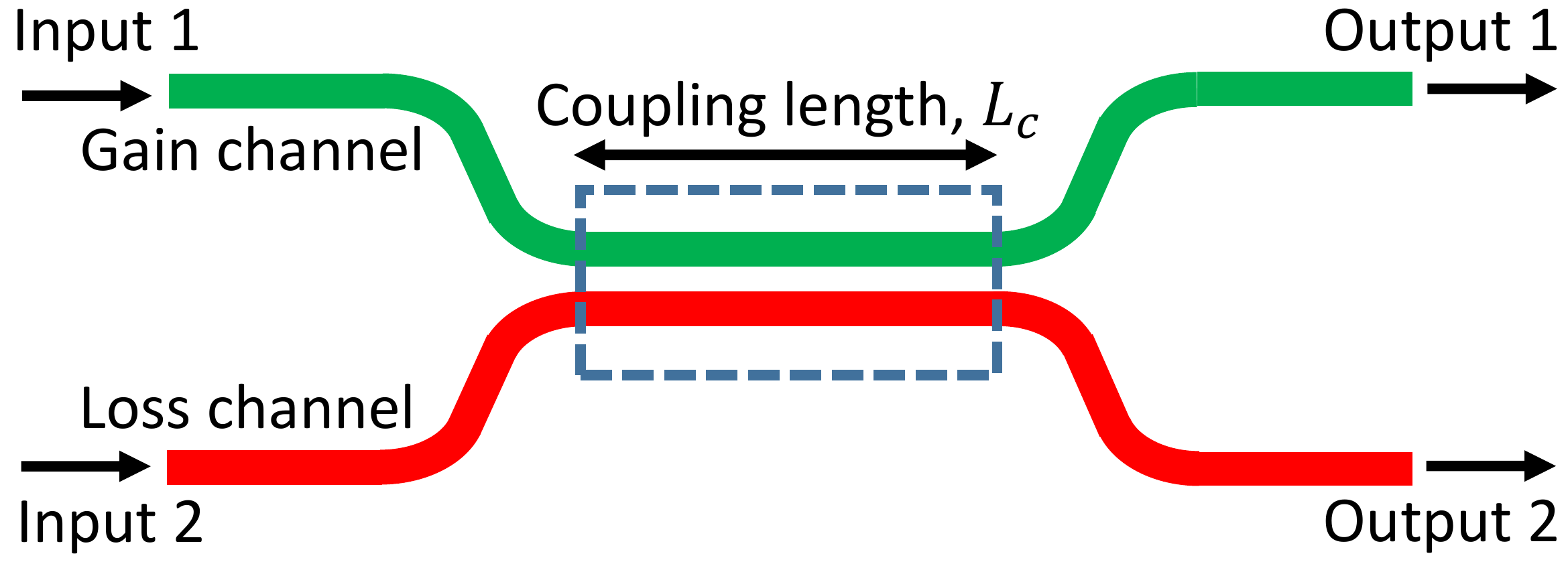}
\caption{(Color online) Schematic diagram of a $\PT$-symmetric nonlinear directional fiber coupler.}\label{Fig-1} \vspace{-0.5cm}
\end{center}
\end{figure}
%

\vspace{-0.2cm}
\section{Model: $\PT$-symmetric nonlinear fiber coupler} \label{Sec2}
\vspace{-0.2cm}
In order to describe the soliton switching dynamics in a realistic $\PT$-symmetric fiber coupler with Kerr nonlinearity [a schematic diagram of which is depicted in Fig.\,\ref{Fig-1}], we consider a generalized NLSE for individual channel of the coupler with higher-order dispersion (HOD) [third-order dispersion (TOD) and higher] and higher-order nonlinear terms [intrapulse Raman scattering (IRS) and self-steepening (SS)] acting as perturbations \cite{GPAbook1}. The dimensionless coupled-mode equation of the slowly-varying field envelops $\psi_{1,2}(\xi,\tau)$ in the two respective channels $(1,2)$ of the coupler can be written in the long pulse regime (where the integral form of nonlinear term follows the derivative form \cite{GPAbook1}) as \cite{GPAbook2, Govindarajan19}:
\begin{subequations}\label{eq1} \vspace{-0.4cm}
\begin{align}
i\,{\partial_\xi \psi_1} +& \sum_{n=2}^\infty \delta_n \left(i\,\partial_\tau\right)^n \psi_1 +|\psi_1|^2 \psi_1+i\,s\,\partial_\tau\left(|\psi_1|^2\psi_1 \right) \nonumber \\ 
&\hspace{1.2cm}  -\tau_R\,\psi_1\,\partial_\tau\left(|\psi_1|^2\right) +\kappa \,\psi_2 =  i\,\Gamma\, \psi_1, \label{eq1a} \\
i\,{\partial_\xi \psi_2} +& \sum_{n=2}^\infty \delta_n \left(i\,\partial_\tau\right)^n \psi_2 +|\psi_2|^2 \psi_2+i\,s\,\partial_\tau\left(|\psi_2|^2\psi_2 \right) \nonumber \\
&\hspace{0.9cm}-\tau_{R}\,\psi_2\,\partial_\tau\left(|\psi_2|^2\right)+\kappa \,\psi_1= -i\,\Gamma\, \psi_2.  \label{eq1b}
\end{align}
\end{subequations}
Here, $\xi$, $\tau$, $\kappa$, $\Gamma$, $\delta_n$, $\tau_R$, and $s$ are the dimensionless length, time, linear coupling, balanced gain/loss, group-velocity dispersion (GVD) ($n=2$) and HOD ($n>2$), IRS, and SS, respectively (rescaled by the characteristic dispersion length of the medium, and the input peak power and pulse duration \cite{GPAbook1}).

The $\pm\Gamma$ terms in Eqs.\,\eqref{eq1a} and \eqref{eq1b} refer to the gain in the first channel and loss in the second channel of the coupler with equal magnitude. This balanced gain and loss makes the coupler a $\PT$-symmetric one, and for which the non-Hermitian optics suggests that there exists three operational domains: unbroken ($\Gamma<\kappa$), broken ($\Gamma>\kappa$), and an exceptional point (EP) ($\gamma=\kappa$) \cite{El-Ganainy18}. In this work, however, we restrict our analysis to the unbroken regime as we investigate the switching dynamics between two channels that are primarily confined in the unbroken regime \cite{Govindarajan19, Sahoo22}. When we approach the singularity point ($\Gamma=\kappa$), the pulse evolution dynamics becomes unstable. Furthermore, the variational method fails to capture the pulse dynamics close to the EP.

Note that, when all the higher-order perturbations are absent i.e., ($\delta_{n,\, n>2}=0$, $s=0$, and $\tau_R=0$) the set of coupled equations [Eqs.\eqref{eq1a} and \eqref{eq1b}] describes the unperturbed case of a $\mathcal{PT}$-symmetric coupler for $\Gamma\neq 0$ \cite{Govindarajan19} and its conventional counterpart for $\Gamma=0$ \cite{GPAbook2}. The numerical simulations are carried out using split-step fast-Fourier transform technique complemented with Runge-Kutta algorithm by launching  $\sqrt{P_0}\,{\rm sech}(\tau)$ soliton solution (which is an exact solution of the unperturbed single NLSE) at the input, where $P_0$ is the dimensionless input peak power related to the square of the soliton order. Before proceeding with the detailed numerical analysis, in the next section (Sec.\,\ref{Sec3}), we develop a detailed analysis of Lagrange's variational approach for $\PT$-symmetric couplers in the context of soliton switching dynamics considering higher-order perturbation terms. The variational method is employed with the above mentioned $sech$ ansatz function, as the success of this method is dependent on the proper choice of the ansatz function appropriate for the system under investigation.

\vspace{-0.2cm}
\section{Variational analysis} \label{Sec3}
\vspace{-0.2cm}
To apply the variational method, we first write the coupled-mode equations [Eqs.\eqref{eq1a} and \eqref{eq1b}] in the form of perturbed coupled NLSEs \cite{GPAbook2}
\begin{align} \label{eq2}
i{\partial_\xi \psi_{1,2}} + \frac{1}{2}{\partial_\tau^2 \psi_{1,2}} +|\psi_{1,2}|^2\psi_{1,2} + \kappa \psi_{2,1}=i\epsilon_{1,2},
\end{align}
where the dispersion is taken to be anomalous ($\delta_2 =-1$)
and define the nonconservative and higher-order perturbation terms through $\epsilon_{1,2}$ as
\begin{align} \label{eq3}
\epsilon_{1}= \Gamma\,\psi_1 +{\epsilon_h}_1, ~~\epsilon_{2}= -\Gamma\,\psi_2 +{\epsilon_h}_2,
\end{align}
where, ${\epsilon_h}_{1,2}=\delta_3\,\partial_\tau^3\psi_{1,2}-s\,\partial_\tau\left(|\psi_{1,2}|^2\psi_{1,2} \right)-i\,\tau_{R}\,\psi_{1,2}\,\partial_\tau\left(|\psi_{1,2}|^2\right)$.
Note that, here we have taken the TOD as the only HOD term in the perturbed NLSE to make the variational method calculation simple. Also, this TOD perturbation contributes significantly to the switching dynamics compared to the next HOD terms.
Now, introducing a Lagrangian density $\left(\mathcal{L_D} \right)$ appropriate for Eq.\,\eqref{eq2}
\begin{align} \label{eq4}
\mathcal{L_D}=&\sum_{j=1,2}\left\{-{\rm Re}[i\psi_j\partial_\xi\psi_j^*]+({1}/{2})\left(|\psi_j|^4    -|\partial_\tau \psi_j|^2 \right) \right. \nonumber \\&\hspace{2.2cm} \left.-2{\rm Re}[i\epsilon_j\psi_j^*]  \right\}  +2\kappa\,{\rm Re}[\psi_1^*\psi_2],
\end{align}
and considering mathematical form of ansatze for Kerr solitons
\begin{align} \label{eq5}
\psi_{j}(\xi,\tau)=\sqrt{\frac{{E_{j}(\xi)\,\eta_j(\xi)}}{2}}{\rm sech}[\eta_j(\xi)\{\tau-\tau_p(\xi)\}] \times \nonumber \\
\times \exp[i\,\phi_{j}(\xi)-i\,\Omega_p(\xi)\{\tau-\tau_p(\xi)\}],
\end{align}
we obtain a reduced Lagrangian $\left(L = \int_{-\infty}^{\infty}\mathcal{L_D}\, d\tau \right)$ of the following form:
\begin{align} \label{eq6}
&\hspace{-0.05cm}L = \sum_{j=1,2}\left\{\frac{E_j}{6}\left(E_j \eta_j -\eta_j^2 -3\Omega_p^2\right)- E_j\left({\partial_\xi\phi_j} +\Omega_p \, {\partial_\xi\tau_p} \right)  \right. \nonumber \\ &\hspace{1.0cm}\left. -2\,{\rm Re}\int_{-\infty}^{\infty}\left[i\epsilon_j \psi_{j}^* \right]d\tau \right\} +2\kappa\sqrt{E_1 E_2}\, \cos\Phi. 
\end{align}
Here $\Phi=\phi_1 - \phi_2$, and the eight ansatze parameters $E_{1,2}$ (pulse energy), $\eta_{1,2}$ (inverse of temporal pulse width),  $\phi_{1,2}$ (phase), $\tau_p$ (peak temporal position), and $\Omega_p$ (peak spectral position) are  assumed to evolve with $\xi$. 
It is worth noting that the ansatze $\psi_j$ considered in Eq.\,\eqref{eq5} have same $\tau_p$ and $\Omega_p$ for both channels of the coupler, as both the pulses experience identical perturbations and thus give equal amounts of temporal and spectral shifts (which can be further verified numerically). In order to evaluate the integration associated with the coupling $\kappa$ [the last term in Eq.\eqref{eq4}] and to get closed form of the integration [the last term in Eq.\eqref{eq6}], we consider the above mentioned conditions (i.e., same $\tau_p$ and $\Omega_p$ between two channels) along with $\eta_1=\eta_2$. Also, we have not considered the frequency chirp parameter in the above ansatze [Eq.\,\eqref{eq5}] to make the total calculations a much simpler, else solving the set of coupled equations is extremely difficult. Despite these sacrifices, our variational results excellently predict the spatiotemporal soliton evolution and switching dynamics that we are aiming for. Before that, we proceed with the next step of the variational method where we use Euler-Lagrange equation for each ansatz parameter to obtain a set of coupled ordinary differential equations describing overall spatio-temporal soliton dynamics. These equations are as follows:
\begin{align}
&\frac{d E_1}{d\xi}=2\kappa \sqrt{E_1 E_2}\,\sin\Phi +2\,{\rm Re}\int_{-\infty}^{\infty}\epsilon_1 \, \psi_1^* \,d\tau, \label{eq7} \\
&\frac{d E_2}{d\xi}=-2\kappa \sqrt{E_1 E_2}\,\sin\Phi +2\,{\rm Re}\int_{-\infty}^{\infty}\epsilon_2 \, \psi_2^* \,d\tau, \\
&\frac{d \tau_p}{d\xi}=-\Omega_p + 2\,{\rm Re}\int_{-\infty}^{\infty}\frac{(\tau-\tau_p) }{E_1+E_2}\left[\epsilon_1 \, \psi_1^* +\epsilon_2 \, \psi_2^* \right]d\tau, \label{eq9}\\
&\frac{d \Omega_p}{d\xi}=-\frac{1}{E_1+E_2}{\rm Im}\int_{-\infty}^{\infty}\left[\mathcal{F}_1\, \epsilon_1 \, \psi_1^*  +\mathcal{F}_2\,\epsilon_2 \, \psi_2^* \right]d\tau,  \label{eq10} \\
&\frac{d \Phi}{d\xi}= \frac{1}{3}\left(E_1\eta_1 -E_2\eta_2 \right)                       -\frac{1}{6}\left(\eta_1^2 -\eta_2^2 \right)-\kappa\frac{E_1-E_2}{\sqrt{E_1 E_2}}\cos\Phi  \nonumber \\
&\hspace{1.8cm}+{\rm Im}\int_{-\infty}^{\infty}\left[\frac{1}{E_1}\epsilon_1 \, \psi_1^* -\frac{1}{E_2}\epsilon_2 \, \psi_2^* \right]d\tau, \\
&\eta_{j, \,j=1,2}= \frac{E_j}{2} +\frac{6}{E_j}{\rm Im}\int_{-\infty}^{\infty} \mathcal{G}_j \, \epsilon_j \, \psi_j^*\, d\tau, \label{eq12}
\end{align}
where $\mathcal{F}_j=2\eta_j \tanh\{\eta_j(\tau-\tau_p)\}$ and $\mathcal{G}_j=\left[1 - (\tau-\tau_p)\mathcal{F}_j\right]/(2\eta_j)$.
The final step is to evaluate all the integrals of Eqs.\,\eqref{eq7}-\eqref{eq12} using $\epsilon_{1,2}$, which results in the evolution of individual pulse parameters. In the following sections (Secs.\,\ref{Sec4} and \ref{Sec5}), we consider different possible coupler scenario for studying pulse evolution and switching dynamics both numerically and using variational method.

\vspace{-0.2cm}
\section{$\mathcal{PT}$-symmetric coupler: the unperturbed case} \label{Sec4}
\begin{figure}[t]
\centering
\begin{center}
\includegraphics[width=0.49\textwidth]{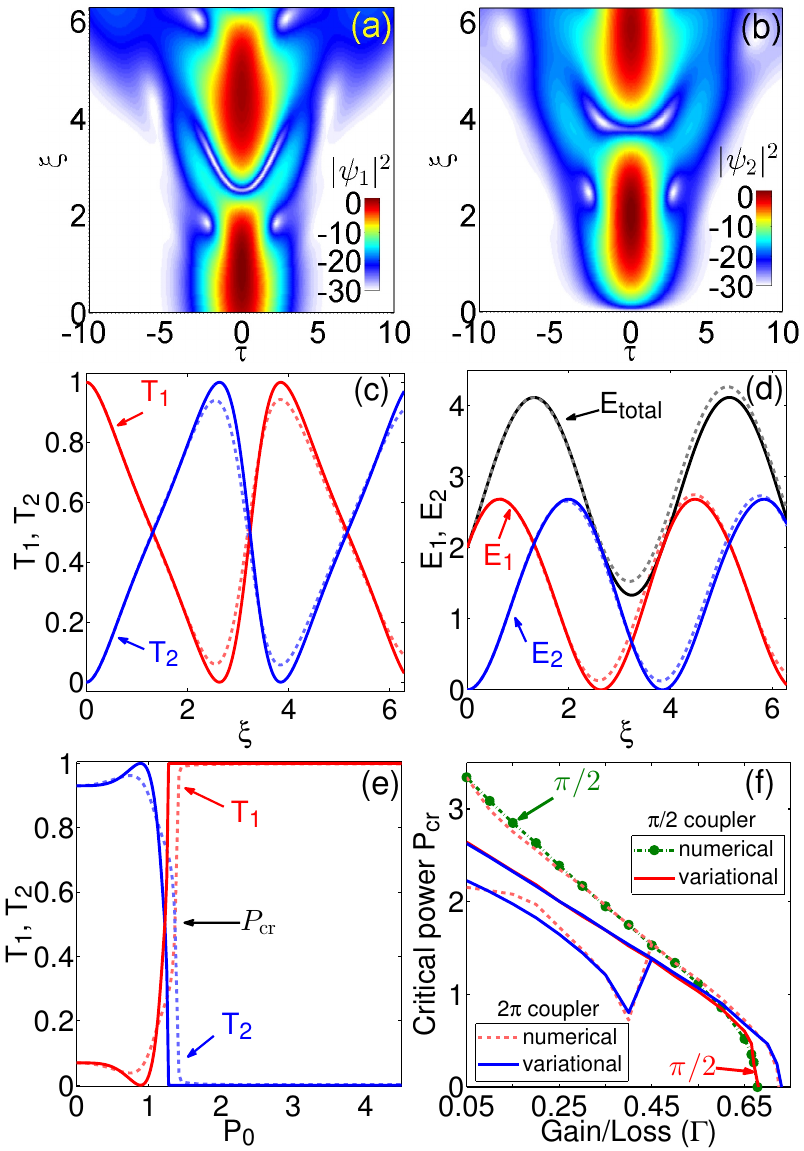}
\caption{(Color online) Switching dynamics for $\mathcal{PT}$-symmetric unperturbed couplers. (a),(b) Spatiotemporal evolution of solitons (in dB) and  their (c) transmission efficiencies ($T_{1,2}$) and (d) energies ($E_{1,2}$) for $\kappa=1$, $\Gamma=0.5$, and $P_0=1$. (e) Switching characteristics ($T_{1,2}$ vs $P_0$) for a $2\pi$ coupler and (f) $P_{\rm cr}$ vs $\Gamma$ for both $2\pi$ and $\pi/2$ coupler configurations.
Solid curves in (c)-(f) represent the variational results, while light dashed curve represent the numerical results. Here, $P_0=E_1(\xi=0)/2$ for plotting the variational results. The coupled equations [Eqs.\,\eqref{eq13}-\eqref{eq15}] are solved with initial conditions (at $\xi=0$): $E_1=2$, $E_2=10^{-6}$ and $\Phi=0$ for (b) and (c); $E_2=10^{-6}$ and $\Phi=0$ for (e) and (f) with varying $E_1$.}\label{Fig-2} \vspace{-0.5cm}
\end{center}
\end{figure}

\vspace{-0.2cm}
In this section, we investigate the soliton evolution and switching dynamics in detail, considering an unperturbed case of the $\PT$-symmetric coupler. The unperturbed situation can be practically realized in the picosecond time-scale (higher-order nonlinear effects are negligible and can be ignored) by launching the input pulse far away from the zero-GVD wavelength (HOD terms can be neglected). Neglecting the ${\delta_n}_{,\,n\ge 3},~s,$ and $\tau_R$ terms from Eq.\,\eqref{eq1} and solving with $\psi_1={\rm sech}(\tau)$ and $\psi_2=0$, we plot the evolution in two channels of a $2\pi$ $\PT$-coupler (coupling length, $L_c=2\pi/\kappa$) in Figs.\,\ref{Fig-2}(a,b). In Figs.\,\ref{Fig-2}(c,d), we also plot the transmission efficiency $T_{1,2}[=P_{1,2}/(P_1 + P_2)$ with $P_{1,2}=\int_{-\infty}^\infty |\psi_{1,2}(L_c,\tau)|^2 d\tau]$ and total integral power of the pulse (pulse energy) $E_{1,2}(=\int_{-\infty}^\infty |\psi_{1,2}(\tau)|^2 d\tau)$ in two channels (light-dashed curves), which show back and forth energy oscillations. This also demonstates that $T_1$ and $T_2$ are $\pi$ phase apart, whereas $E_1$ and $E_2$ are $\pi/2$ phase apart, a characteristic feature common to $\PT$-symmetric couplers \cite{Govindarajan19}. 
%
%

For the variational analysis, here we evaluate the integrals of the set of equations [Eqs.\,\eqref{eq7}-\eqref{eq12}] with $\epsilon_1=\Gamma\,\psi_1$ and $\epsilon_2=-\Gamma\,\psi_2$, achieving the following set of equations describing the evolution dynamics of ansatze parameters:
\begin{align}
\frac{d E_1}{d\xi}&=2\kappa \sqrt{E_1 E_2}\,\sin\Phi +2\,\Gamma \,E_1, \label{eq13} \\
\frac{d E_2}{d\xi}&=-2\kappa \sqrt{E_1 E_2}\,\sin\Phi -2\,\Gamma\,E_2,  \label{eq14} \\
\frac{d \Phi}{d \xi}&=\frac{1}{8}\left(E_1^2 -E_2^2 \right) -\kappa\frac{\left(E_1 - E_2 \right)}{\sqrt{E_1E_2}} \cos\Phi. \label{eq15}
\end{align}
%
\begin{figure}[t]
\centering
\begin{center}
\includegraphics[width=0.49\textwidth]{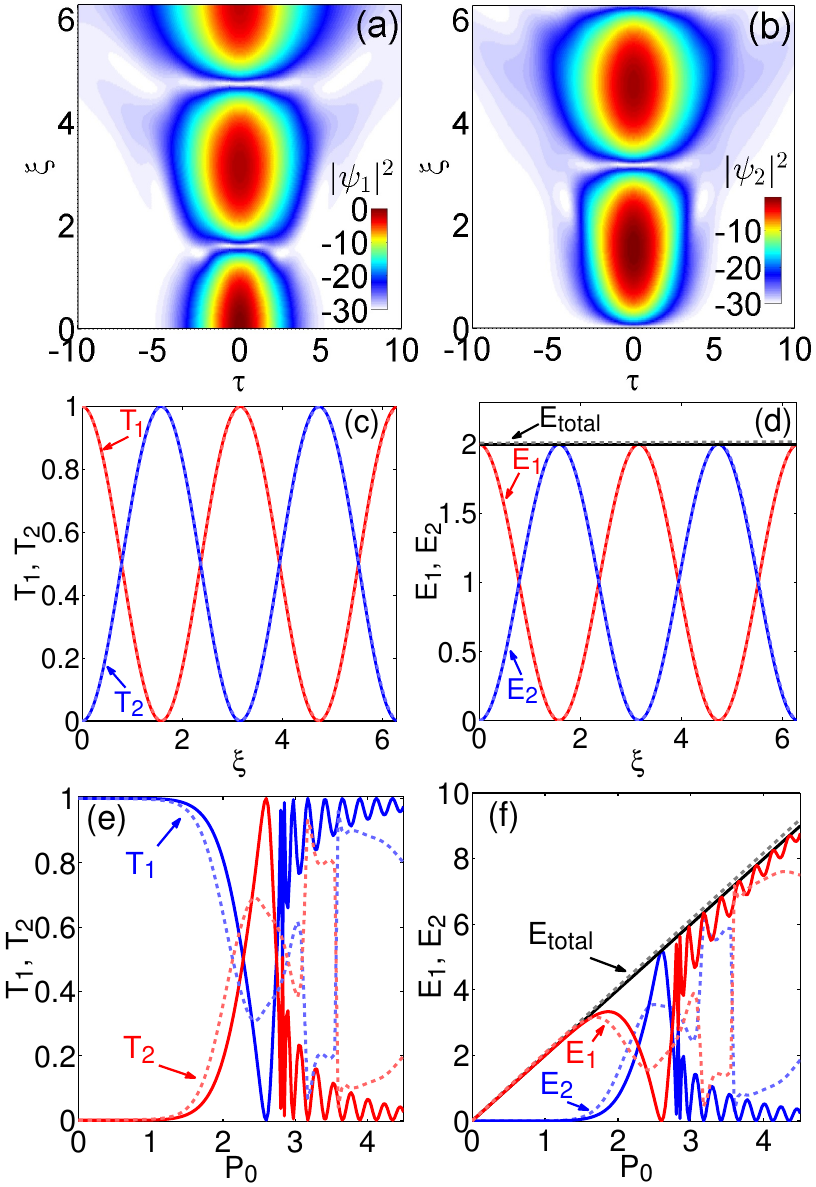}
\caption{(Color online) Switching dynamics for conventional unperturbed couplers. (a)-(f) All the plots represent the same as Fig.\,\ref{Fig-2} with $\Gamma=0$. The other parameters and initial conditions are the same as in Fig.\,\ref{Fig-2}.}\label{Fig-3} \vspace{-0.5cm}
\end{center}
\end{figure}
%
This set of equations [Eqs.\,\eqref{eq13}-\eqref{eq15}] enables one to get physical insights into the unperturbed $\mathcal{PT}$-symmetric couplers. For example, the $\Gamma$ term appears in both the energy equations [Eqs.\,\eqref{eq13} and \eqref{eq14}] with $+$ and $-$ signs corresponding to gain and loss, while the coupling and phase terms ($\kappa\sin\Phi$) associated with these two equations imply back and forth energy oscillations between two channels. Next, to analytically capture the switching dynamics we solve these set of coupled equations [Eqs.\,\eqref{eq13}-\eqref{eq15}] and compare the results with the full numerical simulations of Eq.\,\eqref{eq1}. The findings are quite interesting. The variational results exactly predict the numerical findings of these energy oscillations as depicted by solid curves in Figs.\,\ref{Fig-2}(c,d). Note that the equations for $\tau_p$ [Eq.\,\eqref{eq9}] and $\Omega_p$ [Eq.\,\eqref{eq10}] do not appear in the set of equations [Eqs.\,\eqref{eq13}-\eqref{eq15}] according to the initial launching conditions. In Fig.\,\ref{Fig-2}(e), we plot the transmission efficiency at the output of a $2\pi$ $\PT$-symmetric coupler as a function of input power, which shows a sharp transmission with $~99\%$ energy transfer. Next, in order to get a complete picture of the switching characteristics of the $\PT$-symmetric coupler, we plot the critical power $P_{\rm cr}$ (where $T_1=T_2=0.5$) as a function of $\Gamma$ in Fig.\,\ref{Fig-2}(f) for both $2\pi$ and $\pi/2$ (coupling length, $L_c=\pi/2\kappa$) coupler configurations. Here, below $\Gamma\approx 0.45$, $2\pi$ $\PT$-coupler shows two $P_{\rm cr}$, indicating that with the increase of input power the pulse steers twice and eventually exits from the same port. However, as $\Gamma$ approaches the singularity point (beyond $\Gamma\approx 0.4$), the $P_{\rm cr}$ becomes single-valued and gradually decreases to zero (at $\Gamma\approx 0.7$). In contrast, $\pi/2$ $\PT$-coupler exhibits a single $P_{\rm cr}$ line over the range of $\Gamma$, a unique feature of the $\PT$-symmetric couplers. The variational method also predicts the same switching dynamics in both $2\pi$ and $\pi/2$ cases (solid curves). However, there is a slight difference between the numerical results and variational predictions, which can be attributed to the fact that we use the same pulse width while integrating the coupling term. Moreover, the pulse breaks to generate side lobes at the region of minimum intensity [as seen in Figs.\,\ref{Fig-2}(a,b)], which might have contributed to further mismatch. Nevertheless, the overall predictions are reasonably good, as the analytical method predicts bistable and monostable $P_{\rm cr}$ values. It is worth noting that the variational method can predict dynamics even in the nonlinear switching domains ($P_0> P_{\rm cr}$), as shown in Fig.\,\ref{Fig-2}(e,f), which predicts switching dynamics for higher-order input solitons ($P_0>1$).

Next, to keep the study in perspective we consider the case of conventional coupler \cite{GPAbook2} by setting $\Gamma=0$ in Eq.\,\eqref{eq1}. The numerical simulations are performed using the same ansatze functions as with $\PT$-symmetric case and evolutions are plotted in Figs.\,\ref{Fig-3}(a,b). In this case, the output power emerges from the same channel of the $2\pi$ coupler as opposed to its $\PT$-counterpart. The $T_{1,2}$ and $E_{1,2}$ both are out of phase [shown in Figs. \,\ref{Fig-3}(c,d)]. This phase difference in $E_{1,2}$ can also be evident from Eqs.\,\eqref{eq13} and \eqref{eq14}, where the back and forth energy oscillations between the two channels are $\pi$ phase apart when $\Gamma=0$. However, $\Gamma$ introduces an additional phase, as shown in the $\PT$-symmetric case. The numerical (light-dashed curves) and variational predictions (solid curves) excellently match each other (both curves overlap). The multi-steering nature of the $T_{1,2}$ and $E_{1,2}$ are also numerically plotted in Figs.\,\ref{Fig-3}(e,f) (light-dashed curves) for the $2\pi$-coupler, which is further predicted  by the variational method. 

All of the preceding analyses clearly demonstrate that $\PT$-symmetric couplers outperform the conventional one in terms of switching efficiency, sharp switching with stable and and ultra-low $P_{\rm cr}$. It is also to be noted that the $2\pi$ $\PT$-symmetric Kerr coupler is superior than the $\pi/2$ $\PT$-symmetric Kerr coupler in terms of efficiency and lower $P_{\rm cr}$ when used as all optical switching devices. As a result, when analyzing higher-order perturbations in the following section (Sec.\,\ref{Sec5}), we only consider $2\pi$ $\PT$-symmetric couplers.

\vspace{-0.2cm}
\section{$\mathcal{PT}$-symmetric couplers with higher-order perturbations} \label{Sec5}
\vspace{-0.2cm}
Higher-order perturbations (TOD, IRS, and SS) appear in realistic silica fibers when femtosecond (fs) pulse dynamics are concerned. The origin of these physical perturbations and their effects on ultrashort fs pulses have been thoroughly investigated over the years \cite{GPAbook1}. The TOD perturbation leads to the generation of dispersive radiation across a zero-GVD wavelength \cite{Akhmediev95}. IRS causes the self-frequency redshift of the pulse-spectrum \cite{Doran88}, giving rise to a temporal deceleration of the propagating pulse. The SS effect, on the other hand, induces asymmetric spectral broadening by creating optical shock at the leading edge of the temporal pulse \cite{GPAbook1}. In conventional couplers, the impact of IRS has been investigated in detail \cite{Malomed97}. Also, it has recently been numerically demonstrated that the combined effects of higher-order perturbations stabilize the soliton evolution in $\PT$-symmetric couplers at higher gain/loss values \cite{Sahoo21}.
%
\begin{figure}[t]
\centering
\begin{center}
\includegraphics[width=0.49\textwidth]{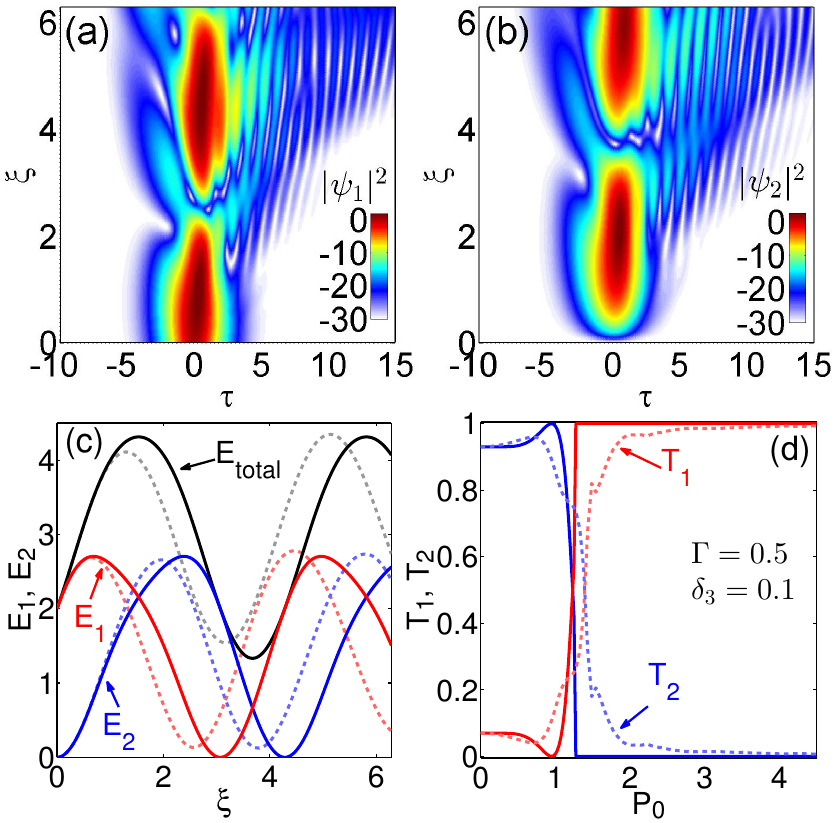}
\caption{(Color online) Switching dynamics for $\mathcal{PT}$-symmetric coupler with TOD perturbation only ($\delta_3 =0.1$). (a),(b) Spatiotemporal evolution of solitons in two channels and (c) their energies for $\kappa=1$, $\Gamma=0.5$, and $P_0=1$. (d) Switching characteristics ($T_{1,2}$ vs $P_0$) for a $2\pi$ coupler. (c),(d) Solid curves represent the variational predictions, while light dashed curve represent the numerical results. Here, the coupled equations [Eqs.\,\eqref{eq16}-\eqref{eq21}] are solved for TOD alone with initial conditions (at $\xi=0$): $E_1=2$, $E_2=10^{-6}$, and $\tau_p=\Omega_p=\Phi=0$ for (c); $E_2=10^{-6}$ and $\tau_p=\Omega_p=\Phi=0$ for (d) with varying $E_1$.}\label{Fig-4} \vspace{-0.5cm}
\end{center}
\end{figure}
%
%
In this section, we theoretically investigate the switching dynamics in the presence of higher-order perturbations using the variational analysis. For that, we evaluate the integrals of the set of coupled equations [Eqs.\,\eqref{eq7}-\eqref{eq12}] with the perturbations [Eq.\,\eqref{eq3}], yielding the final set of coupled equations describing the evolution of ansatze parameters in the presence of perturbations:
\begin{align}
&\frac{d E_1}{d\xi}=2\kappa \sqrt{E_1 E_2}\,\sin\Phi +2\,\Gamma \,E_1, \label{eq16} 
\end{align}
\begin{align}
&\frac{d E_2}{d\xi}=-2\kappa \sqrt{E_1 E_2}\,\sin\Phi -2\,\Gamma\,E_2,  \label{eq17} \\
&\frac{d \tau_p}{d\xi}=-\Omega_p + \frac{s \left(\eta_1 E_1^2 +\eta_2 E_2^2 \right) + 2\,\delta_3\left(\mathcal{H}_1 +\mathcal{H}_2 \right) }{2\,(E_1+E_2)}, \label{eq18} \\
&\frac{d \Omega_p}{d\xi}=-\frac{4}{15}\tau_R \frac{E_1^2 \eta_1^3 +E_2^2 \eta_2^3}{E_1+E_2}, \label{eq19} \\
&\frac{d \Phi}{d\xi} = -\frac{1}{6}\left(\eta_1^2 -\eta_2^2 \right)   -\kappa\frac{E_1-E_2}{\sqrt{E_1 E_2}}\cos\Phi  +\nonumber \\ 
& \hspace{0.8cm}+\frac{1}{3}(1+s\,\Omega_p)\left(E_1\eta_1 -E_2\eta_2 \right) +\delta_3(\mathcal{I}_1 -\mathcal{I}_2), \label{eq20}\\
&\eta_{j, \,j=1,2}= \frac{E_j}{2} +\frac{1}{2}s\,\Omega_p E_j +6\delta_3\Omega_p \eta_j, \label{eq21}
\end{align}
where $\mathcal{H}_j=(\eta_j^2 +3\Omega_p^2)E_j$ and $\mathcal{I}_j=(2-\Omega_p)\eta_j^2$. This set of equations [Eqs.\,\eqref{eq16}-\eqref{eq21}] describes which of the specific perturbation impacts which of the ansatze parameters. For example, $\tau_R$ appears in the $\Omega_p$ equation [Eq.\,\eqref{eq19}] with a negative sign, implying frequency redshifting. This frequency redshifting leads to temporal acceleration, as can be seen from the $\tau_p$ equation [Eq.\,\eqref{eq18}] through the term $-\Omega_p$. The impacts of other perturbations ($\delta_3$ and $s$) are also directly visible from the set of equations [Eqs.\,\eqref{eq16}-\eqref{eq21}].
%
\begin{figure}[t]
\centering
\begin{center}
\includegraphics[width=0.49\textwidth]{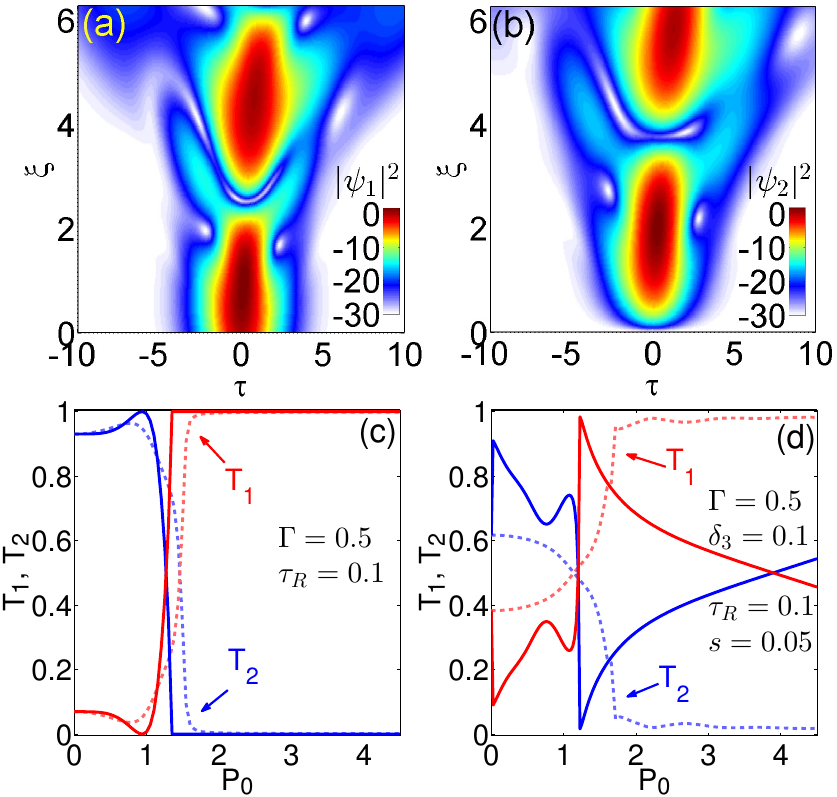}
\caption{(Color online) Switching dynamics for  $\mathcal{PT}$-symmetric coupler with IRS perturbation only ($\tau_R =0.1$). (a),(b) Solito evolution in two channels for $\kappa=1$, $\Gamma=0.5$, $P_0=1$, and (c) the switching characteristics for a $2\pi$ coupler setup. (d) Switching characteristics considering combined effects of all perturbations ($\delta_3=0.1$, $\tau_R=0.1$, and $s=0.05$). Here, solid curves represent the variational results, while light dashed curve represent the numerical results. The coupled equations [Eqs.\,\eqref{eq16}-\eqref{eq21}] are solved with initial conditions (at $\xi=0$): $E_2=10^{-6}$ and $\tau_p=\Omega_p=\Phi=0$ for (c) with varying $E_1$ (IRS alone); $E_2=10^{-6}$, $\eta_1=1$, and $\eta_2=\tau_p=\Omega_p=\Phi=0$ for (d) with varying $E_1$ (TOD, IRS, and SS).}\label{Fig-5} \vspace{-0.5cm}
\end{center}
\end{figure}
%

To capture the pulse evolution and switching dynamics, we first numerically solve the Eq.\,\eqref{eq1} while taking individual perturbations into account. The combined effects of overall perturbations are then subsequently investigated. In Figs.\,\ref{Fig-4}(a,b), we plot the pulse evolution in two channels of the $\PT$-symmetric coupler in the case of TOD alone ($\delta_3\neq 0$, $\tau_R=0$, and $s=0$) considering $\psi_1={\rm sech}(\tau)$ and $\psi_2=0$ as inputs. Here, the generation of characteristic dispersive radiations is evident as side lobes on the right side of the main pulses in two channels. The back and forth energy oscillations ($E_{1,2}$) are plotted in Fig.\,\ref{Fig-4}(c) by light-dashed curves. In Fig.\,\ref{Fig-4}(d), the switching efficiencies $T_{1,2}$ are plotted by light-dashed curves, illustrating that TOD perturbation alone degrades switching efficiency compared to the unperturbed case. In all cases, the variational method (solid curves) qualitatively predicts the pulse evolution and switching efficiency. The deviation between these two occurs due to the fact that resonant radiation can not be captured using the variational method because the main pulse breaks to form temporal side lobes (this also breaks one of the assumptions of the variational method). Next, we numerically confirm that the SS perturbation alone cannot alter the switching dynamics significantly. This effect is taken into account when the combined effects of all perturbations are considered. Using the same input ansatze as before, we perform numerical simulations of Eq.\,\eqref{eq1} while considering IRS as the only perturbation ($\tau_R\neq 0$). The pulse evolutions are depicted in Figs.\,\ref{Fig-5}(a,b), which demonstrate characteristic IRS-induced temporal deceleration. The switching efficiencies are also plotted numerically in Fig.\,\ref{Fig-5}(c) (light-dashed curves). Here, the variational method (solid curves) also accurately predicts the switching dynamics.

As a final test, we run the simulation considering all the perturbations ($\delta_3,~\tau_R,~s$ are all $\neq 0$). The switching efficiencies are plotted in Fig.\,\ref{Fig-5}(d) by light-dashed lines. Here, although the variational method (solid curves) accurately predicts the $P_{\rm cr}$ value, it fails to predict the overall switching characteristics at higher $P_0$ values. This deviation could be exacerbated by the pulse breaking of higher-order solitons in the presence of combined perturbations. Furthermore, we have not taken into account the frequency chirp, which plays an important role in soliton dynamics in the presence of higher-order perturbations. Finally, looking back at all of the cases discussed above, we can see that the variational method is quite effective in predicting the soliton evolution and switching dynamics in $\PT$-symmetric couplers.

\vspace{-0.2cm}
\section{Conclusions} \label{Sec6}
\vspace{-0.2cm}
In conclusion, we have carried out a detailed variational study of the coupled nonlinear Schr\"{o}dinger equation in the context of a nonlinear $\mathcal{PT}$-symmetric coupler. We have addressed the issue of, technologically relevant, soliton steering and dynamics in a $\mathcal{PT}$-symmetric coupler. It is found that numerical calculations are at par with variational calculations, thereby putting the analysis on a strong footing. It is demonstrated that $\PT$-symmetric couplers outclass conventional couplers in terms of switching efficiency, sharp switching with stable and ultra-low critical power. It is found that the so-called $2\pi$ $\PT$-symmetric Kerr coupler is better than the $\pi/2$ $\PT$-symmetric Kerr coupler in terms of efficiency and lower critical power when used as all-optical switching devices. The excellent agreement between the variational calculations and numerical simulations should convince both experimentalists and theorists alike to explore the nonlinear $\PT$-symmetric Kerr coupler further for various applications, including the ones related to solitonic optical communications.

\section*{acknowledgments}
A.S. acknowledges the Ministry of Education, India for a research fellowship through IPDF IIT Guwahati. A.K.S. acknowledges support from the Science and Engineering Research Board (SERB), under the project MATRICS (Grant No. MTR/2019/000945).

\end{document}